\documentclass[10pt]{emulateapj}
\usepackage{apjfonts}
\usepackage{amsmath}
\usepackage{rotating}
\usepackage{enumerate}
\usepackage{natbib}
\newcommand{\pivec}{\mbox{\boldmath $\pi$}}

\shorttitle{}
\shortauthors{}

\begin{document}

\title{
Two Jupiter-Mass Planets Discovered by the KMTNet Survey in 2017
}

\author{
I.-G.~Shin\altaffilmark{H1,K1},
Y.-H.~Ryu\altaffilmark{K1},
J.~C.~Yee\altaffilmark{H1},
A.~Gould\altaffilmark{K1,E1,E2}, 
M.~D.~Albrow\altaffilmark{K3},
S.-J.~Chung\altaffilmark{K1,K2},
C.~Han\altaffilmark{K5},
K.-H.~Hwang\altaffilmark{K1},
Y.~K.~Jung\altaffilmark{K1},
Y.~Shvartzvald\altaffilmark{E3}
W.~Zang\altaffilmark{E4},
C.-U.~Lee\altaffilmark{K1,K2},  
S.-M.~Cha\altaffilmark{K1,K4},   
D.-J.~Kim\altaffilmark{K1},
H.-W.~Kim\altaffilmark{K1,K2},
S.-L.~Kim\altaffilmark{K1,K2},
Y.~Lee\altaffilmark{K1,K4},
D.-J.~Lee\altaffilmark{K1},
B.-G.~Park\altaffilmark{K1,K2},
R.~W.~Pogge\altaffilmark{E1}\\
}

\bigskip\bigskip
\affil{$^{H1}$Harvard-Smithsonian Center for Astrophysics, 60 Garden St., Cambridge, MA 02138, USA}
\affil{$^{K1}$Korea Astronomy and Space Science Institute, 776 Daedeokdae-ro, Yuseong-Gu, Daejeon 34055, Republic of Korea}
\affil{$^{K2}$Korea University of Science and Technology, 217 Gajeong-ro, Yuseong-gu, Daejeon 34113, Republic of Korea}
\affil{$^{K3}$University of Canterbury, Department of Physics and Astronomy, Private Bag 4800, Christchurch 8020, New Zealand}
\affil{$^{K4}$School of Space Research, Kyung Hee University, Giheung-gu, Yongin, Gyeonggi-do, 17104, Republic of Korea}
\affil{$^{K5}$Department of Physics, Chungbuk National University, Cheongju 28644, Republic of Korea}
\affil{$^{E1}$Department of Astronomy, Ohio State University, 140 W. 18th Ave., Columbus, OH 43210, USA}
\affil{$^{E2}$Max-Planck-Institute for Astronomy, K\"onigstuhl 17, 69117 Heidelberg, Germany}
\affil{$^{E3}$IPAC, Mail Code 100-22, Caltech, 1200 E. California Blvd., Pasadena, CA 91125, USA}
\affil{$^{E4}$Physics Department and Tsinghua Centre for Astrophysics, Tsinghua University, Beijing 100084, People's Republic of China}

\begin{abstract}
We report two microlensing events, KMT-2017-BLG-1038 and KMT-2017-BLG-1146 that are caused by planetary systems. These events were discovered by KMTNet survey observations from the $2017$ bulge season. The discovered systems consist of a planet and host star with mass ratios, $5.3_{-0.4}^{+0.2} \times 10^{-3}$ and $2.0_{-0.1}^{+0.6} \times 10^{-3}$, respectively. Based on a Bayesian analysis assuming a Galactic model without stellar remnant hosts, we find that the planet, KMT-2017-BLG-1038Lb, is a super Jupiter-mass planet ($M_{\rm p}= 2.04_{-1.15}^{+2.02}\,M_{\rm J}$) orbiting a mid-M dwarf host ($M_{\rm h}= 0.37_{-0.20}^{+0.36}\, M_{\odot}$) that is located at $6.01_{-1.72}^{+1.27}$ kpc toward the Galactic bulge. The other planet, KMT-2017-BLG-1146Lb, is a sub Jupiter-mass planet ($M_{\rm p}= 0.71_{-0.42}^{+0.80}\,M_{\rm J}$) orbiting a mid-M dwarf host ($M_{\rm h}= 0.33_{-0.20}^{+0.36}\,M_{\odot}$) at a distance toward the Galactic bulge of $6.50_{-2.00}^{+1.38}$ kpc. Both are potentially gaseous planets that are beyond their hosts' snow lines. These typical microlensing planets will be routinely discovered by second-generation microlensing surveys, rapidly increasing the number of detections. 
\end{abstract}

\keywords{gravitational lensing: micro -- exoplanets}

\section{Introduction}

 Extrasolar planets have been discovered and confirmed using several different techniques such as radial velocity (RV), transit, direct imaging, microlensing, etc. In recent years, the total number of detections has rapidly increased, reaching over three thousand ($\simeq 3791$ as of September $27$, $2018$), and includes $\simeq629$ multiple planet systems\footnote[1]{The number counts based on confirmed planets from the NASA Exoplanet Archive (https://exoplanetarchive.ipac.caltech.edu, as of September $27$, $2018$)}. Each of these various methods uses different physical processes for detecting and characterizing planets. This implies that each method is sensitive to a certain category of planets according to their physical properties. For example, RV is most sensitive to planets located inside the snow lines of their hosts, while microlensing is sensitive to planets located beyond the snow line. Thus, these methods are complementary in providing observational constraints for understanding planets.

 Studies of the planet frequency around M dwarf hosts using different surveys, i.e., RV \citep{johnson10, bonfils13} and microlensing \citep{gould10}, yield conflicting results \citep[see Figure 9 of][]{gould10}. The microlensing surveys \citep{gould10,cassan12} found that giant planets beyond the snow line are more common than those probed by RV surveys. Thus, \citet{clanton14a,clanton14b} state that ``it is not clear if this is a consequence of a lack of formation or lack of migration'' but also question whether or not the results can, in fact, be reconciled. They investigated the overall planet demographics by combining the results of the RV and microlensing detection methods. They found that the demographic constraints inferred from the two methods are actually consistent based on their analysis. However, at that time, their study was conducted based on relatively limited microlensing samples. In particular, they used a total of $23$ microlensing planets from \citet{gould10} and \citet{sumi10}.

 In the era of second-generation microlensing surveys, the number of microlensing planet detections is rapidly increasing. As detections increase (not only by microlensing but also by the other methods), this can provide an opportunity to independently compare the planet frequencies inferred by several methods for the overlapping region of planet parameter space. Although the synthesis of planet demographics can in principle describe our current knowledge of planets orbiting M dwarf hosts \citep{clanton16}, the comparison of independent studies could provide more direct and clear observational constraints.

 Among the microlensing surveys that can lead to routine planet detections is the Korea Microlensing Telescope Network \citep[KMTNet:][]{kim16}. In particular, KMTNet was designed to near-continuously monitor wide fields toward the Galactic bulge with high-cadence observations, which are sufficient to catch planetary anomalies on the light curves induced by planets with masses ranging from (super) Jupiters down to Earths. Indeed, since the $2015$ commissioning season, the KMTNet survey has provided observations that are crucial constraints to detect and characterize several planets including planet candidates by catching planetary perturbations on microlensing light curves \citep[e.g.,][]{albrow18, calchi18a, calchi18b, han16, han17a, han17b, hwang18a, hwang18b, jung18a, miyazaki18, mroz17, mroz18, ryu18, shin16, shvartzvald17, skowron18, zang18}. Based on the observations of the commissioning season, KMTNet established its own microlensing event finder \citep{kim18}. Thus, KMTNet successfully achieved the independent detection of its first planet candidate: KMT-2016-BLG-0212 \citep{hwang18a}.

 To date, the total number of microlensing planet detections has reached $64$ (as of September 27, 2018). In particular, $\sim 47\%$ of planets discovered by the microlensing are giant planets \citep[$M_{\rm p}/M_{\rm Jupiter} \ge 0.1$ adopting the definition of giant planet from][]{clanton14b} that are located beyond the snow line and orbiting M dwarf hosts ($0.08 < M_{\rm h}/M_{\odot} < 0.6$). Considering the detection sensitivity of the microlensing method, such giant planets are quite typical and will be routinely discovered.

 During the $2017$ bulge season, the KMTNet survey was able to detect several KMT-only planets/planet candidates \citep[e.g.,][]{jung18b}. Among them, we report here on two typical microlensing planets: a super Jupiter-mass planet and on a sub Jupiter-mass planet, which are named KMT-2017-BLG-1038Lb and KMT-2017-BLG-1146Lb, respectively.

 In this paper, we describe the observations and light curves of these planets in Section 2. In Section 3, we present the analysis of these light curves. In Section 4, we characterize the properties of the discovered planetary systems using Bayesian analyses. Lastly, we discuss our findings in Section 5.

\section{KMTNet Observations}

 KMTNet consists of three identical $1.6$ m telescopes with wide-field ($4\,{\rm deg^2}$) cameras, which are located in the Cerro Tololo Inter-American Observatory in Chile (KMTC), the South African Astronomical Observatory in South Africa (KMTS), and the Siding Spring Observatory in Australia (KMTA). The observatory locations on three continents of the southern hemisphere allow near-continuous observations covering wide fields toward the Galactic bulge. 

 The high-cadence observation strategy of the KMTNet survey aims to discover and characterize planets without additional follow-up observations. To achieve this purpose, the required cadence of observations ranges from $\Gamma \ge 4\,{\rm hr^{-1}}$ to $>0.2\,{\rm hr^{-1}}$ for discoveries ranging from Earth-mass to Jupiter-mass planets, respectively. These cadences derive from the short duration of the planetary signals $t_p \sim t_{\rm E} \sqrt{q} \sim 5(q/10^{-4})^{1/2}$ hr, where the $q$ is the mass ratio of the planet and host star and $t_{\rm E}$ is the Einstein timescale of the microlensing event (typical timescale is $\sim20$ days). The KMTNet survey monitors a wide field to maximize the number of events. It monitors $12\,{\rm deg^2}$, $41\,{\rm deg^2}$, $85\,{\rm deg^2}$, and $97\,{\rm deg^2}$ with observation cadences $\Gamma \sim 4\,{\rm hr^{-1}}$, $>1\,{\rm hr^{-1}}$, $>0.4\,{\rm hr^{-1}}$, and $>0.2\,{\rm hr^{-1}}$, which are sufficient to detect and characterize Earth, Neptune, Saturn, and Jupiter class planets, respectively.

\begin{figure}[htb!]
\epsscale{1.00}
\plotone{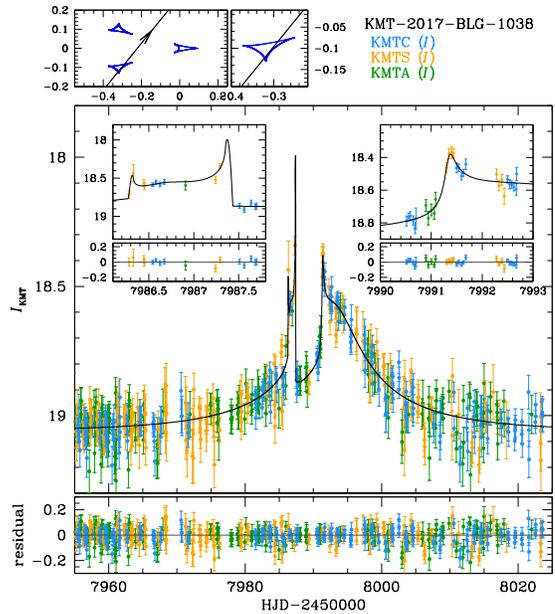}
\caption{Light curve of KMT-2017-BLG-1038. The upper two panels show the geometry of the event with a zoom-in of the transit of the region enclosed by one of the two triangular planetary caustics. The lower panels show the observed light curve and the best-fit model curve with residuals. The inset panels show zoom-ins of caustic transit (left) and cusp approach (right) parts, respectively. Each colored dot indicates KMTNet observations taken from each observatory: CTIO (light blue), SAAO (yellow), and SSO (green).
\label{fig:lc1038}}
\end{figure}

 The observations are mainly made with the {\it I-}band channel. In addition, KMTNet regularly takes {\it V-}band images. In $2017$, {\it V-}band images were roughly $10\%$ of the total observations taken from KMTC and $5\%$ from KMTS and KMTA. These {\it V-}band data can be used for the construction and analysis of the color-magnitude diagrams (CMD) to extract color information of the lensed star. These data sets were reduced by the KMTNet pipeline, which employs the image subtraction method \citep[pySIS:][]{albrow09}. These light curves were run through the KMTNet event finder \citep{kim18} to find microlensing events. In brief, this event finder finds microlensing events by fitting observed light curves (simultaneously fitted all data taken from three KMT sites) on the single-lensing light curve using an efficient 2D grid of ($t_0$, $t_{\rm eff}$), rather than a prohibitive 3D grid of ($t_0$, $u_0$, $t_{\rm E}$)\footnote[2]{For detailed formalism and descriptions, see Section 2 and 3 of \citet{kim18}.}, where $t_{\rm eff}\rightarrow{u_{0}t_{\rm E}}$. From this machine review, microlensing event candidates were produced. These candidates were manually reviewed to classify them as clear/possible microlensing events, artifacts, or several classes of variables. The events in this work were found by this process (classified as clear microlensing events).

\subsection{Observations of KMT-2017-BLG-1038}

 The microlensing event, KMT-2017-BLG-1038, occurred on a background star (source) located at $(\alpha,\delta)_{\rm J2000}=(17^{h}44^{m}41^{s}.02,-25^{\circ}08^{'}34^{''}.91)$ corresponding to the Galactic coordinates $(l,b)=(3.^{\circ}13,2.^{\circ}16)$. This event was discovered by the KMTNet survey alone.

 In Figure \ref{fig:lc1038}, we present the observed light curve of this event. The light curve exhibits dramatic deviations from the single-lensing light curve. The anomaly has complex features caused by a transit of the regions enclosed by one triangular planetary caustic (from ${\rm HJD'}\simeq7986.3$ to ${\rm HJD'}\simeq7987.4$) and an approach to a cusp of the other planetary caustic (${\rm HJD'}\simeq7991.4$). These features are covered by KMTC and KMTS observations.

\begin{figure}[htb!]
\epsscale{1.00}
\plotone{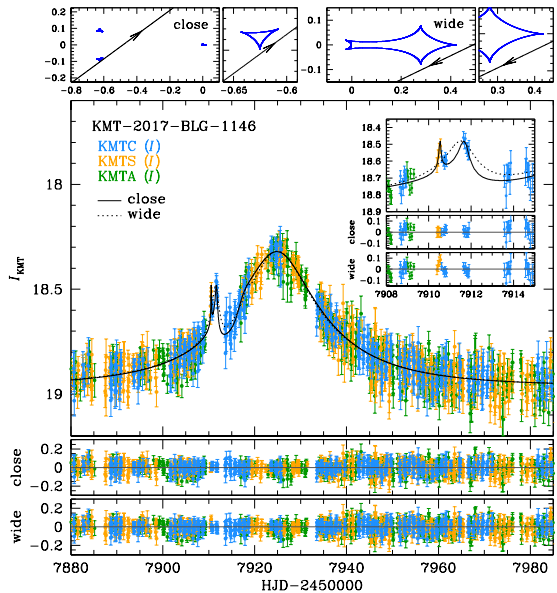}
\caption{Light curve of KMT-2017-BLG-1146. The upper panels show the geometries of the close and wide solutions with zoom-ins of caustic-approaching parts, respectively. The lower panels show the observed light curve and both model curves of the close (dash line) and wide (dotted line) solutions with their residuals. The inset panel shows a zoom-in of the caustic-approaching part with residuals of the close and wide solutions. The color scheme of the observations is identical to Figure \ref{fig:lc1038}.
\label{fig:lc1146}}
\end{figure}

 We note that the data have relatively large uncertainties due to the faintness of the source, i.e., $I_{\rm KMT}\sim21.4$ mag at the baseline (unmagnified source brightness). In addition, the caustic entrance (${\rm HJD'}\simeq7986.3$) and exit (${\rm HJD'}\simeq7986.4$) are not optimally covered. Thus, these gaps in coverage cause large uncertainties of model parameters (see Section 3.2). For the {\it V-}band data, due to the faint source, the signal-to-noise of the {\it V-}band data is extremely low. Thus, it was not possible to extract the color information of the source from these data.

\subsection{Observations of KMT-2017-BLG-1146}

 The microlensing event, KMT-2017-BLG-1146, occurred on a source that lies at $(\alpha,\delta)_{\rm J2000}=(17^{h}56^{m}25^{s}.40,$\\$-33^{\circ}08^{'}32^{''}.89)$ corresponding to the Galactic coordinates $(l,b)=(-2.^{\circ}44,-4.^{\circ}14)$. This event was also discovered by the KMTNet alone.

 In Figure \ref{fig:lc1146}, we present the observed light curve of KMT-2017-BLG-1146, which has a typical planetary anomaly (${\rm HJD'}\simeq7911.0$) on the underlying single-lensing light curve. The anomaly is caused by the caustic approach of the source. The KMTC and KMTS observations precisely covered this anomaly, which plays a key role in distinguishing between degenerate models (see Section 3.3 for the details). 

 However, for the {\it V-}band observations, the data are too uncertain to determine the source color, which is similar to the case of KMT-2017-BLG-1038.

\section{Light Curve Analysis}

\subsection{Modeling Process}

 To build models for describing the observed light curves, we adopt a standard parameterization that consists of seven parameters: $t_0, u_0, t_{\rm E}, s, q, \alpha,$ and $\rho_{\ast}$, where $t_0$ is the time of the closest approach of lens-source, $u_0$ is the closest separation (normalized to the Einstein radius) between the source and a reference position on the binary-axis at the time of $t_0$ (i.e., impact factor), $t_{\rm E}$ is the timescale to cross the Einstein radius, $s$ is a separation between binary-lens components normalized by the Einstein radius, $q$ is the mass ratio of the binary-lens components, $\alpha$ is the angle with respect to the binary-axis, and $\rho_{\ast}$ is the angular source radius normalized by the Einstein radius. This parameterization is described in \citet{jung15} with a conceptual figure. Based on this parameterization, we conduct modeling processes that consist of two steps: a grid search followed by refining the model.

\begin{deluxetable}{lrrr}
\tablecaption{Best-fit Model Parameters and Error Rescaling Factors\label{table:one}}
\tablewidth{0pt}
\tablehead{
\multicolumn{1}{c}{event} &
\multicolumn{1}{c}{KMT-2017-BLG-1038} & 
\multicolumn{2}{c}{KMT-2017-BLG-1146} 
}
\startdata
$\chi^2 / {\rm N_{data}}$  & $\bf{2075.774 / 2076}$            & $\bf{1945.248 / 1945}$            & $1954.702 / 1945$            \\
$t_0$                      & $\bf{7992.829_{-0.071}^{+0.088}}$ & $\bf{7924.890_{-0.132}^{+0.035}}$ & $7924.787_{-0.143}^{+0.035}$ \\
$u_0$                      & $\bf{   0.172_{-0.006}^{+0.007}}$ & $\bf{   0.287_{-0.017}^{+0.032}}$ & $   0.213_{-0.006}^{+0.036}$ \\
$t_{\rm E}$                & $\bf{  21.902_{-0.6}^{+0.7}}$     & $\bf{  25.443_{-1.8}^{+1.2}}$     & $  30.587_{-3.2}^{+0.8}$     \\
$s$                        & $\bf{   0.851_{-0.003}^{+0.003}}$ & $\bf{   0.734_{-0.017}^{+0.010}}$ & $   1.148_{-0.008}^{+0.020}$ \\
$q$ $(\times 10^{-3})$     & $\bf{   5.3_{-0.4}^{+0.2}}$       & $\bf{   2.0_{-0.1}^{+0.6}}$       & $   4.5_{-0.4}^{+1.5}$       \\
$\alpha$                   & $\bf{   5.396_{-0.019}^{+0.018}}$ & $\bf{   5.649_{-0.036}^{+0.005}}$ & $   2.693_{-0.024}^{+0.002}$ \\
$\rho_{\ast, {\rm limit}}$ & $\bf{   < 0.004                }$ & $\bf{   < 0.010                }$ & $   < 0.033                $ \\
$\rho_{\ast, {\rm best}}$  & $\bf{     0.0012               }$ & $\bf{     0.0004               }$ & $     0.0007               $ \\
$F_{\rm S,KMTC}$           & $\bf{   0.042_{-0.001}^{+0.002}}$ & $\bf{   0.127_{-0.009}^{+0.018}}$ & $   0.090_{-0.003}^{+0.018}$ \\
$F_{\rm B,KMTC}$           & $\bf{   0.334_{-0.002}^{+0.001}}$ & $\bf{   0.286_{-0.019}^{+0.008}}$ & $   0.322_{-0.018}^{+0.002}$ \\
\hline                                                                                    
KMTC                       & \bf{1.314}                        & \bf{1.218}                        & 1.218                        \\
KMTS                       & \bf{1.500}                        & \bf{1.305}                        & 1.305                        \\
KMTA                       & \bf{1.528}                        & \bf{1.150}                        & 1.150                        \\
\enddata
\tablecomments{
The uncertainties reflect the true significant digits. However, more decimal places 
for the values are provided so that the interested reader may reproduce the best-fit model.
The $\rho_{\ast, {\rm limit}}$ present $3\sigma$ upper limits (see Figure \ref{fig:dist1038} and \ref{fig:dist1146}).
The $\rho_{\ast, {\rm best}}$ present the best-fit values that used to draw the light curves in Figure \ref{fig:lc1038} and \ref{fig:lc1146}.
We present these $\rho_{\ast}$ values for readers who may be interested to reproduce the best-fit models.
}
\end{deluxetable}

 The first process is a grid search. We perform a grid search with dense grid sets, i.e., $[s, q]$ to find global and local minima. The grid parameters are chosen to be ($s$, $q$, and $\alpha$) because $(s, q)$ are directly related to the caustic geometry, and the various source trajectories defined by the $\alpha$ parameter yield dramatic changes in the light curve features for a fixed caustic. The grid parameters have ranges $\log(s) = (-1.0, 1.0)$ and $\log(q) = (-5.0, 1.0)$, which cover almost all cases of lensing lightcurves caused by various caustic geometries from binary-lens systems down to planet-host systems. The ranges of the grid parameters are densely divided into $n(\log(s))=100$ and $n(\log(q))=100$. For each $(s, q)$ grid point, a total of $21$ values of the $\alpha$ parameter are used to seed solutions from $\alpha=(0^{\circ}, 360^{\circ})$ along with the other four parameters, i.e., $t_0$, $u_0$, $t_{\rm E}$, and $\rho_{\ast}$. These five parameters are allowed to vary continuously from the seed solutions. Thus, the $\alpha$ parameter is a kind of semi-grid parameter: a grid is used to seed the initial value of $\alpha$, but then it is allowed to vary to find the optimal solution. In the grid search process, to find a global or local minimum, we use a $\chi^2$ minimization method adopting a Markov Chain Monte Carlo (MCMC) algorithm \citep{dunkley05}.

\begin{figure}[htb!]
\epsscale{1.00}
\plotone{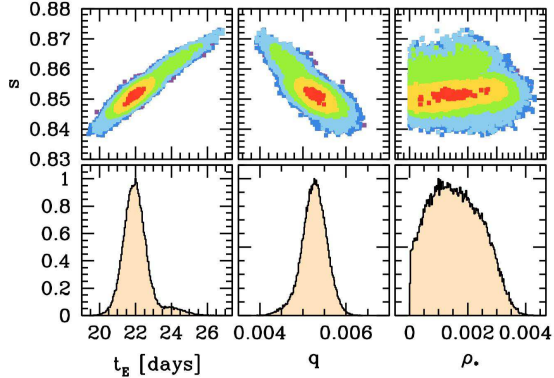}
\caption{MCMC chain scatters and distributions of KMT-2017-BLG-1038. In the upper three panels, we present the MCMC chain scatters for selected model parameters, $t_{\rm E}$, $q$, and $\rho_{\ast}$, with respect to the parameter $s$. Each color represents $\Delta\chi^2$ from the best-fit model: $1^{2}$ (red), $2^{2}$ (yellow), $3^{2}$ (green), $4^{2}$ (sky blue), $5^{2}$ (blue), and $6^{2}$ (purple). The lower three panels show the distributions of selected parameters normalized by each peak value.
\label{fig:dist1038}}
\end{figure}

 The second modeling process is refining the model parameters. For each global or local minimum, we refine the model parameters, which are able to vary all parameters within all possible ranges using the MCMC sampling method. During this process, one data point should on average give $\Delta\chi^2\sim1$. Thus, we rescale the errors using an equation, $e_{\rm new} = k e_{\rm old}$, where $e_{\rm new}$ are the rescaled errors, $e_{\rm old}$ are the original errors reported by the reduction software and $k$ is the rescaling factor for each data set presented in Table \ref{table:one}. From this refining process, we can estimate the uncertainties of the model parameters based on the MCMC chain. The uncertainties are determined based on the $68\%$ confidence intervals around parameters of the best-fit model.

\subsection{Model of KMT-2017-BLG-1038}

 In Table \ref{table:one}, we present the best-fit model parameters and error rescaling factors of KMT-2017-BLG-1038. From the modeling process, we find that there exists only one global minimum with $q\sim0.005$ and $s\sim0.851$. The observed light curve of this event shows complex features produced by a transit of the region enclosed by one of the two triangular, planetary caustics (from ${\rm HJD'}\sim7986.3$ to ${\rm HJD'}\sim7987.5$) and a cusp approach (${\rm HJD'}\sim7991.4$). Moreover, there exists a region of reduced magnification between these two features. Thus, these complex perturbations on the light curve can be described by the special caustic geometry of a $s<1$ (close) case, i.e., the source crosses the lower planetary caustic and then approaches the upper planetary caustic. In between these planetary caustics, the source traverses a negative magnification region (relative to a point lens). Because $s>1$ (wide) solutions cannot produce such large regions of negative magnification, the close/wide degeneracy \citep{griest98,dominik99}, which is a well-known degeneracy that can prevent the unique determination of the binary-lens properties (especially a planetary lensing event), is decisively broken for this event.

\begin{figure}[htb!]
\epsscale{1.00}
\plotone{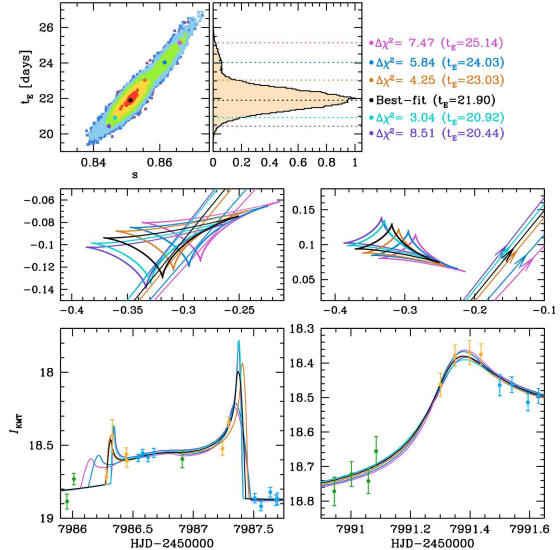}
\caption{Various model curves of KMT-2017-BLG-1038 within $3\sigma$ of the best fit. The upper two panels show the scatters and distributions of the ($t_{\rm E}$) parameter. The color scheme of the scatters is identical to those of Figure \ref{fig:dist1038}. Each colored dot indicates extracted representative solutions within $3\sigma$ of the best fit including the best-fit model itself. The lower panels show zoom-ins of geometries and model light curves. The lower-left panels show the caustic transit part and the lower-right panels show the cusp approach.
\label{fig:multi1038}}
\end{figure}

 In Figure \ref{fig:dist1038}, we present the MCMC chain distributions of selected parameters ($t_{\rm E}$, $s$, and $q$) that are essential to determining the lens properties when applying a Bayesian analysis. As seen in the plots, the $t_{\rm E}$ distribution is not a normal Gaussian distribution. Moreover, the distribution has a ``tail'' at the $3\sigma$ level, corresponding to alternative solutions describing the observed light curve. In Figure \ref{fig:multi1038}, we present various representative models extracted from the non-Gaussian distribution within $3\sigma$ of the best solution. We find that the solutions in the ``tail'' cannot perfectly describe the two KMTS data points that covered the entrance of the caustic around ${\rm HJD'}\sim 7986.3$. At the same time, these model light curves from the ``tail'' fit the planetary caustic approach around ${\rm HJD'}\sim7991.35$ better than the best-fit model. As a result, the $\chi^2$ difference between the two families of models is relatively small, and this second family of solutions from the ``tail'' cannot be ruled out.

\subsection{Model of KMT-2017-BLG-1146}

 In the case of KMT-2017-BLG-1146, we find that there is a possibility for degenerate solutions caused by the close/wide degeneracy. From the grid search, we find that the lowest $\chi^2$ solution lies in the close ($s<1$) regime. However, there exists a plausible solution in the wide ($s>1$) regime. Thus, we investigate both local minima. From the model refining process, we find that the wide solution cannot perfectly describe the planetary anomaly on the light curve, especially near ${\rm HJD'}\sim7910.5$ by comparison to the close solution (see the zoom-in of Figure \ref{fig:lc1146}). This difference yields $\Delta \chi^2 \sim 10$ between the close and wide solutions, which is a meaningful $\Delta\chi^2$ value considering the relatively small number of data points covering the anomaly. As a result, we conclude that the close solution with $q\sim0.002$ and $s\sim0.734$ is the preferred solution for describing the observed light curve of this event. In Table \ref{table:one}, we present the parameters of the close solution, and also the parameters of the wide solution although it is disfavored.

 In Figure \ref{fig:dist1146}, we also present the MCMC chain scatters and distributions of selected parameters for this event. In contrast to KMT-2017-BLG-1038, the $t_{\rm E}$ distribution of this case follows a normal Gaussian distribution.

\subsection{Higher-order Effects}

\begin{figure}[htb!]
\epsscale{1.00}
\plotone{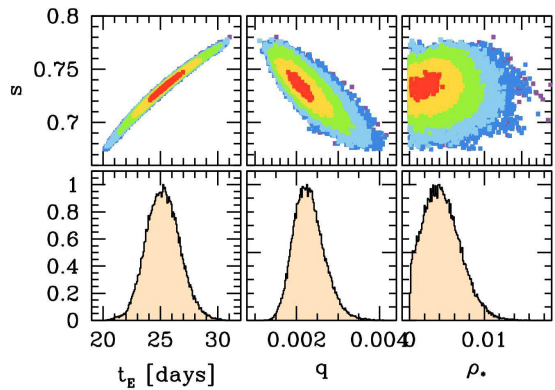}
\caption{MCMC chain scatters and distributions of KMT-2017-BLG-1146. The format is the same as in Figure \ref{fig:dist1038}.
\label{fig:dist1146}}
\end{figure}

 Even though the companions in both events are highly likely to be planets based on mass ratios of $q\sim10^{-3}$, additional observables are required to measure the lens and planet masses, the distance to the lens system and the projected separation between the planet and the host star. These observables are the angular Einstein ring radius ($\theta_{\rm E}$) and the magnitude of the microlens parallax ($\pi_{\rm E}$). The Einstein radius, $\theta_{\rm E}$, can be measured from the model parameter, $\rho_{\ast} \equiv \theta_{\ast} / \theta_{\rm E}$ where $\theta_{\ast}$ is the angular source radius, which can be measured from the analysis of the CMD adopting the method from \citet{yoo04}. The microlens parallax $\pi_{\rm E}$ can be measured by observing the annual microlensing parallax effect \citep{gould92} caused by the orbital motion of Earth\footnote[3]{Indeed, there exists an alternative method to measure the microlens parallax, i.e., the satellite microlens parallax \citep{refsdal66, gould94}, by measuring the offset of light curves seen from space and Earth. However, this method requires simultaneous observations with a space telescope, which was not possible in these cases because the events were discovered after the end of the microlensing season. Thus, in this work, we only consider the APRX for the microlens parallax measurement.}, which would introduce additional model parameters, $\pi_{{\rm E}, N}$ and $\pi_{{\rm E}, E}$, i.e., the north and east components of the microlens parallax vector ($\pivec_{\rm E}$).

 Both observables are necessary to analytically determine the lens properties. However, if it is not possible to measure both, even one of these observables would be a constraint for the Bayesian analysis. Thus, we investigate the possibility of measuring these additional observables for the events.

 In Figure \ref{fig:dist1038} and \ref{fig:dist1146}, we present distributions of the $\rho_{\ast}$ parameter (right panels) of KMT-2017-BLG-1038 and KMT-2017-BLG-1146, respectively. For both events, we find that $\rho_{\ast}$ values are consistent with zero at the $2\sigma$ and $1\sigma$ levels, respectively. Thus, there is no significant detection of the finite source effect. However, there are upper limits. 

 We test a model of the annual microlens parallax (APRX) to check the possibility of detecting the APRX signal on the lensing light curve. We check the $\chi^2$ improvement when the APRX parameters ($\pi_{{\rm E}, N}$ and $\pi_{{\rm E}, E}$) are introduced. For the KMT-2017-BLG-1146 event, we find that the $\chi^2$ improvement is only $\sim 0.6$. This improvement is insignificant, so we cannot claim a detection of APRX. For the KMT-2017-BLG-1038 event, we find that the $\chi^2$ improvement is $\sim 13.4$. We carefully investigate this improvement because the data for this event have some systematics, which may produce a false-positive signal of the APRX. From the investigation, we find that the $\chi^2$ improvement comes from only KMTS data, rather than consistent improvements from all data sets. This is unusual behavior considering the data sets cover the event about equally. If the APRX signal were really present, all data sets should show $\chi^2$ improvements. Hence, we conclude that the $\chi^2$ improvement comes from the unknown systematics of the KMTS data, rather than the APRX effect.

\section{Properties of Discovered Planetary Systems}

\subsection{Bayesian Analysis}

 We characterize the discovered planetary systems based on a Bayesian analysis. For this analysis, we generate a total of $4\times10^{7}$ artificial microlensing events using a Monte Carlo Simulation. For the Galactic priors, we adopt the velocity distributions of \citet{hangould95}, the mass functions of \citet{chabrier03}, and matter density profiles of the Galactic bulge and disk as compiled by \citet{hangould03} (for the details of the Bayesian formalism, see Section 4 of \citet{jung18a} and references therein.). We note that we take into account the line of sight to the actual event when the artificial events are generated. We also note that the host types (i.e., normal stars or stellar remnants) of artificial events are defined when these events are generated. The mass fractions of stellar remnants are calculated by adopting observational constraints of several studies (white dwarfs: \citealt{bragaglia95}, neutron stars: \citealt{thorsett99}, and black holes: \citealt{gould00}).

\begin{deluxetable}{lrr}
\tablecaption{Properties of Discovered Planetary Systems \label{table:two}}
\tablewidth{0pt}
\tablehead{
\multicolumn{1}{c}{} &
\multicolumn{1}{c}{KMT-2017-BLG-1038} & 
\multicolumn{1}{c}{KMT-2017-BLG-1146}
}
\startdata
%
w/ stellar remnants            &                         &                         \\
$M_{\rm host}$ $(M_{\odot})$   & $0.43_{-0.25}^{+0.32}$  & $0.40_{-0.25}^{+0.34}$  \\
$M_{\rm planet}$ $(M_{\rm J})$ & $2.4_{-1.4}^{+1.8}$     & $0.85_{-0.52}^{+0.76}$  \\
$D_{\rm L}$ (kpc)              & $6.1_{-1.6}^{+1.2}$     & $6.6_{-1.9}^{+1.4}$     \\
$a_{\perp}$ (au)               & $1.9_{-0.6}^{+0.6}$     & $1.7_{-0.6}^{+0.6}$     \\
$a_{\rm snow}$ (au)            & $1.2_{-0.7}^{+0.9}$     & $1.1_{-0.7}^{+0.9}$     \\
$\mu$ (mas ${\rm yr^{-1}}$)    & $6.3_{-1.9}^{+2.3}$     & $5.3_{-1.9}^{+2.3}$     \\   
\hline                                                                              
w/o stellar remnants           &                         &                         \\ 
$M_{\rm host}$ $(M_{\odot})$   & $0.37_{-0.20}^{+0.36}$  & $0.33_{-0.20}^{+0.36}$  \\
$M_{\rm planet}$ $(M_{\rm J})$ & $2.0_{-1.1}^{+2.0}$     & $0.71_{-0.42}^{+0.80}$  \\
$D_{\rm L}$ (kpc)              & $6.0_{-1.7}^{+1.3}$     & $6.5_{-2.0}^{+1.4}$     \\
$a_{\perp}$ (au)               & $1.8_{-0.5}^{+0.6}$     & $1.6_{-0.6}^{+0.6}$     \\
$a_{\rm snow}$ (au)            & $1.0_{-0.6}^{+1.0}$     & $0.9_{-0.5}^{+1.0}$     \\
$\mu$ (mas ${\rm yr^{-1}}$)    & $6.1_{-1.9}^{+2.3}$     & $5.1_{-1.9}^{+2.4}$     \\   
\enddata
\tablecomments{
The uncertainties are determined based on the $68\%$ confidence intervals around 
median values of Bayesian analyses.
}
\end{deluxetable}

 Based on generating events, we construct probability distributions with respect to the lens mass ($M_{\rm L}$), the distance of the lens ($D_{\rm L}$), the projected separation between the planet and host ($a_{\perp}$), and the Einstein timescale ($t_{\rm E}$) of the generated events. Then, we put a constraint on the probability distribution by applying a weight function. The weight function consists of two parts that are derived from the $t_{\rm E}$ and $\rho_{\ast}$ distributions of our event.

 The first part of the weight function is constructed by fitting the $t_{\rm E}$ distribution of the event based on the superposition of Gaussian functions written as
\begin{equation}
W(t_{\rm E}) = \sum_{i=1}^{2} {a_i}\,e^{ {-\frac{1}{2}}\left({\frac{t_{\rm E}-\mu_i}{\sigma_i}}\right)^2 },
\end{equation}
where the set of coefficients, $(a, \mu, \sigma)$ are determined by fitting of the $t_{\rm E}$ distribution. In case of KMT-2017-BLG-1038, the coefficient sets are determined as $(a_{1}, \mu_{1}, \sigma_{1}) = (1.000, 21.923, 0.582)$ and $(a_{2}, \mu_{2}, \sigma_{2}) = (0.058, 24.089, 0.638)$. In case of KMT-2017-BLG-1146, the coefficient sets are determined as $(a_{1}, \mu_{1}, \sigma_{1}) = (0.097, 26.556, 1.465)$ and $(a_{2}, \mu_{2}, \sigma_{2}) = (0.943, 25.096, 0.943)$. We use this fitting method to reflect the actual $t_{\rm E}$ distribution instead of a normal Gaussian weight because of the non-Gaussian $t_{\rm E}$ distribution of the KMT-2017-BLG-1038 case. Although the $t_{\rm E}$ distribution for KMT-2017-BLG-1146 is close to a normal Gaussian, for consistency in the Bayesian analysis, we apply the identical methodology for the analysis of both events.

\begin{figure}[htb!]
\epsscale{1.00}
\plotone{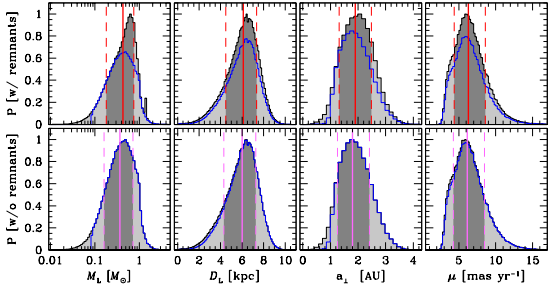}
\caption{Posterior distributions of the Bayesian analyses for KMT-2017-BLG-1038. The upper four panels show posteriors with stellar remnants of the host mass ($M_{\rm L}$), distance to the lens system ($D_{\rm L}$), projected separation between the planet and host ($a_{\perp}$), and relative lens-source proper motion ($\mu$), respectively. the lower four panels show posteriors without stellar remnants. The solid line indicates the median of each distribution. The dark grey shade with the dotted line indicates the $68\%$ confidence interval of each distribution, i.e., the $1\sigma$ uncertainty around the median value. The blue lines indicate distributions excluding non-luminous hosts (i.e., brown dwarfs, white dwarfs, neutron stars, and black holes).
\label{fig:bayesian1038}}
\end{figure}

 The second part of the weight function is constructed from the $\rho_{\ast}$ distribution. Although the detections of the finite source effect of both events are not significant, we apply the $\rho_{\ast}$ distribution to the Bayesian analysis as a constraint because the distribution of $\rho_{\ast}$ values and its upper limit can provide a partial constraint. However, the $\rho_{\ast}$ distribution has a non-Gaussian form. Thus, we construct a weight function using $\Delta\chi^2$ of MCMC chains as a function of $\rho_{\ast}$ values following the method in \citet{calchi18b}. This weight function is written as
\begin{equation}
W(\rho_{\ast}) = e^{-\frac{1}{2} \Delta\chi^2(\rho_{\ast})}~~;~~\\
\end{equation}
\begin{equation}
\Delta\chi^2(\rho_{\ast}) \equiv \begin{cases} 
				 0                                                                & \rho_{\ast} \leq \rho_{\ast,{\rm best}} \\
				 \chi^2_{\rm trial}(\rho_{\ast}) - \chi^2_{\rm best}(\rho_{\ast}) & \rho_{\ast} > \rho_{\ast,{\rm best}}
                                 \end{cases} ,
\end{equation}
where the $\chi^2_{\rm trial}(\rho_{\ast})$ is the $\chi^2$ value of each trial fitting, $\chi^2_{\rm best}(\rho_{\ast})$ is the $\chi^2$ value of the best-fit, and $\rho_{\ast,{\rm best}}$ is a $\rho_{\ast}$ value of the best-fit model. For KMT-2017-BLG-1038, the $\rho_{\ast}$ constraint affects the Bayesian analysis by excluding $\sim 8\%$ of artificial events compared to the analysis using the $t_{\rm E}$ constraint only. By contrast, for KMT-2017-BLG-1146, the effect of the $\rho_{\ast}$ constraint is $< 0.2\%$. However, we apply the same methodology for consistency. Therefore, the final weight function is constructed as $W=W(t_{\rm E}) \cdot W(\rho_{\ast})$. 

\begin{figure}[htb!]
\epsscale{1.00}
\plotone{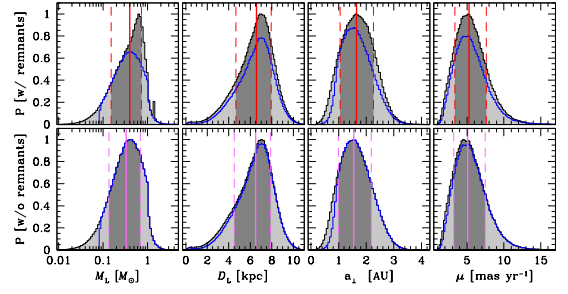}
\caption{Posterior distributions of the Bayesian analyses for KMT-2017-BLG-1146. The description is the same as for Figure \ref{fig:bayesian1038}.
\label{fig:bayesian1146}}
\end{figure}

 Then, we determine properties of the lens system from the weighted probability distribution. In Figures \ref{fig:bayesian1038} and \ref{fig:bayesian1146}, we present the posterior distributions for both KMT-2017-BLG-1038 and KMT-2017-BLG-1146. We note that the properties of planetary systems are determined for cases both with and without consideration of stellar remnants. The chance to discover planets orbiting stellar remnant hosts could be extremely low \citep{kilic09} because the host, e.g., a sun-like star, at the end of its evolution stage (i.e., red giant, asymptotic giant, or planetary nebula) engulfs the planet(s) within $\simeq 1$ au \citep{nordhaus10}. Thus, one version of the Bayesian analysis was done without stellar remnants to reflect a galaxy in which remnants do not host giant planets. However, \citet{mullally08, mullally09} reported a candidate of a gaseous planet ($M \sin{i} \sim 2 \, M_{\rm J}$), which is orbiting a white dwarf star, GD 66. Considering this discovery, we cannot rule out the possibility of planets orbiting a stellar remnant host. Thus, we also consider posterior distributions with stellar remnants as hosts, even though the chance of discovering this kind of planetary system could be very low.

\subsection{Properties of the Planets}

 In Table \ref{table:two}, we present representative properties of the two discovered planetary systems, which are the median and $1\sigma$ uncertainty ($68\%$ confidence interval of the distribution) determined from the posterior distributions. In addition, we estimate the snow line of each planetary system using the relation, $a_{\rm snow} = 2.7\, {\rm au}\, (M_{\rm host}/M_{\odot})$, adopted from \citet{kennedy08}. The planets discovered in KMT-2017-BLG-1038 and KMT-2017-BLG-1146 are super Jupiter-mass and sub Jupiter-mass planets, respectively, both orbiting mid-M dwarf hosts. Both planets are located beyond their own snow lines. 

 We determined the properties of these planetary systems using Bayesian analyses (with or without) stellar remnant hosts. For KMT-2017-BLG-1038Lb, we found that the planet is a super Jupiter-mass planet ($M_{\rm planet} = 2.41_{-1.41}^{+1.80}$ or $2.04_{-1.15}^{+2.02}\, M_{J}$) orbiting a mid-M dwarf host ($M_{\rm host} = 0.43_{-0.25}^{+0.32}$ or $0.37_{-0.20}^{+0.36}\, M_{\odot}$) with projected separation, $1.90_{-0.58}^{+0.59}$ or $1.80_{-0.54}^{+0.61}$ au, which is located beyond the snow line. This system is located at the distance, $6.12_{-1.64}^{+1.23}$ or $6.01_{-1.71}^{+1.27}$ kpc, from us. For KMT-2017-BLG-1146Lb, the planet is a sub Jupiter-mass planet ($M_{\rm planet} = 0.85_{-0.52}^{+0.76}$ or $0.71_{-0.42}^{+0.80}\, M_{J}$) orbiting a mid-M dwarf host ($M_{\rm host} = 0.40_{-0.25}^{+0.34}$ or $0.33_{-0.20}^{+0.36}\, M_{\odot}$) with projected separation, $1.65_{-0.60}^{+0.61}$ or $1.55_{-0.56}^{+0.63}$ au, which is also located beyond the snow line. This system is located at the distance, $6.57_{-1.91}^{+1.36}$ or $6.50_{-2.00}^{+1.38}$ kpc, from us. 

 We found that the relative lens-source proper motions are $\sim 6$ and $\sim 5$ ${\rm mas}\, {\rm yr^{-1}}$ of KMT-2017-BLG-1038 and KMT-2017-BLG-1146, respectively. Ten years after the events, close to the start of thirty meter class telescope operations, the lens and source of these events will be separated with $\ge 60$ and $\ge 50$ mas, respectively. In addition, the estimated brightness of both lenses is $\sim 21$ mag in {\it H}-band. The flux ratios (source/lens) are $\sim 9$ and $\sim 11$ for KMT-2017-BLG-1038 and KMT-2017-BLG-1146, respectively\footnote[4]{The {\it H-}band magnitudes and flux ratios are estimated using median values of the Bayesian analyses.}. Hence, considering the resolving power of the $30$ m-class telescopes ($\theta \sim 14$ mas for {\it H-}band observations), the lenses can be detected in follow-up observations using large telescopes with adaptive optics systems, which will lead to much more precise constraints on the physical properties of the planets (and their hosts). If the host is luminous (see Figure \ref{fig:bayesian1038} and \ref{fig:bayesian1146}), such measurements will yield both a measurement of the lens flux and the lens-source relative proper motion (and hence $\theta_{\rm E}$), thus giving a complete solution for the lens mass and distance \citep[e.g.,][]{yee15}. 

\begin{figure}[htb!]
\epsscale{1.00}
\plotone{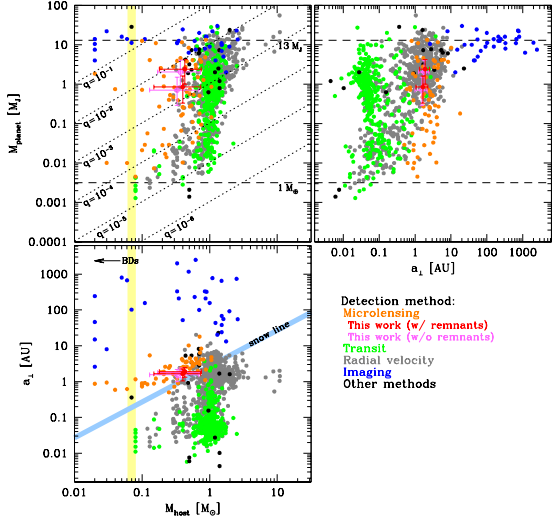}
\caption{Diagrams of confirmed exoplanets with planets of this work. Colored dots indicate confirmed planets detected by a different method. The upper-left panel shows the mass distribution of the host star and planet. The upper-right panel shows the distribution of the planet mass and the semi-major axis or projected separation. The lower panel shows the distribution of the host mass and the semi-major axis or projected separation with the snow line ($a_{\rm snow} \sim 2.7\, {\rm au} (M_{\rm host} / M_{\odot})$) and the conventional mass limit of brown dwarfs ($0.06\sim0.08\, M_{\odot}$). The properties of confirmed planetary systems are adopted from the NASA Exoplanet Archive (https://exoplanetarchive.ipac.caltech.edu).
\label{fig:exo1}}
\end{figure}

\section{Discussion}
 We reported two planets, KMT-2017-BLG-1038Lb and KMT-2017-BLG-1146Lb, discovered by the KMTNet survey in the $2017$ bulge season. In Figure \ref{fig:exo1}, we visualize the physical properties of our discoveries compared to those of other confirmed planets detected by various methods. The microlensing method can detect a wide range of planet masses. Moreover, as is well-known, the method is sensitive to planets that are located beyond the snow line \citep[e.g.,][]{gouldloeb92}. Our findings are typical microlensing planets: these are giant planets beyond the snow line, bound to M-dwarf hosts. Thus, similar planets will be routinely discovered and characterized by the microlensing surveys. As a result, the number of detections will systematically increase in the future.

\begin{figure}[htb!]
\epsscale{0.60}
\plotone{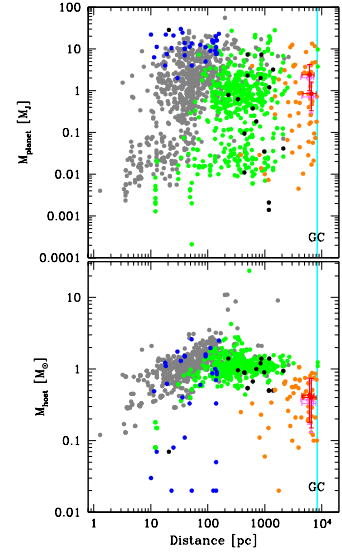}
\caption{Diagrams of the distance dependence of the detection method. The upper panel shows the distribution of planet mass and its distance from us. The lower panel shows the distribution of host mass and its distance from us. Each color of dots (confirmed planetary systems) represents a different detection method shown in Figure \ref{fig:exo1}. The cyan line indicates the distance of the Galactic center.
\label{fig:exo2}}
\end{figure}

 In Figure \ref{fig:exo2}, we also visualize the locations of the discovered planets together with those of the other confirmed planets. This spatial distribution clearly shows the complementary contribution of each planet-detecting method in building a complete sample of exoplanets in our Galaxy. In particular, the microlensing method can cover various types of planets that are located at the farthest distance from us, which would be difficult to detect by the other methods. 

\begin{figure}[htb!]
\epsscale{1.00}
\plotone{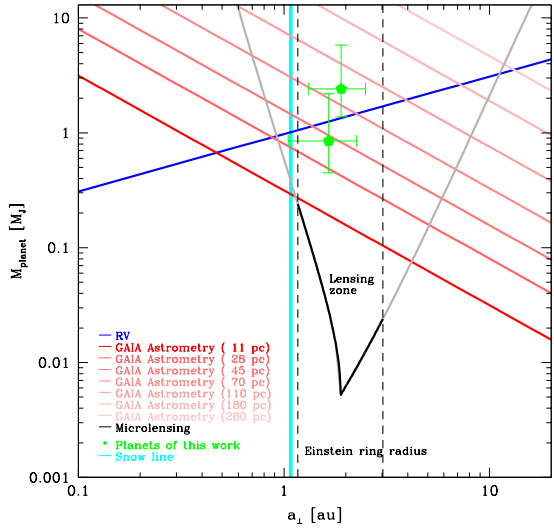}
\caption{
Planet detectabilities of three methods (RV, astrometry, and microlensing). The detectabilities are theoretically estimated based on the physics of each method assuming that the hosts of the planets are mid-M dwarfs ($M_{\rm host} = 0.4\, M_{\odot}$). The blue line indicates the detectability of the RV method assuming that the velocity semi-amplitude ($K$) is $1 {\rm m\, s^{-1}}$. The series of red colors show the detectabilities of the astrometry method adopting a performance of the GAIA telescope for the planetary systems located at various distances from us. The grey line shows the detectability of the microlensing method assuming that the planetary systems are located at $7$ kpc from us. In particular, the black line indicates the theoretical lensing zone \citep{han06}, i.e., with a caustic within the Einstein ring radius (dashed line). The cyan line indicates the snow line of M dwarf host assuming that $a_{\rm snow} = 2.7\, {\rm au}\, (M_{\rm host}/M_{\odot})$ \citep{kennedy08}. The green pentagons show planets discovered in this work. 
\label{fig:detectability}}
\end{figure}

 In Figure \ref{fig:detectability}, we present planet detectabilities of three methods (i.e., RV, astrometry, and microlensing). The detectability is theoretically derived based on the physics of each method assuming planets orbiting a mid-M dwarf host ($0.4\,M_{\odot}$). We present the theoretical detectabilities in the conventional planet parameter space, $a_{\perp}/{\rm au}$ and $M_{\rm p}/M_{\rm Jupiter}$, which are the semi-major axis or projected separation and planet mass, respectively. For the RV detectability, we adopt the analytic equation of \citet{cumming99} for a velocity semi-amplitude of $1\, {\rm m\, s^{-1}}$, which is the assumed performance of a state-of-art RV survey. For astrometry, we adopt a prediction of planet detectability assuming the performance of the GAIA telescope from \citet{perryman14}. In this case, we assume that the planetary systems are located at several distances (from $11$ to $280$ pc) from us. For microlensing, we adopt analytic lensing equations from \citet{gaudi12} and \citet{han06} assuming the caustic must be at least $\Delta\eta_{\rm c}=10^{-2}$ (size relative to the Einstein ring) to be detectable. In this case, we assume that the planetary systems are located near the Galactic bulge ($7$ kpc). This analytic estimate is broadly consistent with the predicted sensitivity of KMTNet \citep{henderson14}. 

 As shown in the Figure, these detectabilities overlap in the region of giant planets ($M_{\rm planet} > 0.1\, M_{\rm Jupiter}$). In addition, these planets approach the snow line from different detections. The growing microlensing planet sample will give us opportunities to independently determine the planet frequency, which can be compared to those of other methods. It will contribute the complementary coverage of the planet parameter space for planet demographics. In the next few years, we can anticipate similar detections of giant planets beyond the snow line orbiting M dwarfs from radial velocities (which will finally have been observing long enough with high enough precision) and GAIA. This will enable a direct comparison of the planet frequency measured from three independent techniques. The results should be consistent, but if they are not, they will reveal some previously unknown systematics in the sampling methods or some variation in planet frequency with Galactic distance (since microlensing primarily probes distant planets whereas RV and GAIA will find those orbiting nearby stars). Such a test is essential for verifying the results from different techniques and testing for the effects of Galactic environment on planet formation. 

\mbox{}

\acknowledgments 
This research has made use of the KMTNet system operated by the Korea Astronomy and Space Science 
Institute (KASI) and the data were obtained at three host sites of CTIO in Chile, SAAO in South 
Africa, and SSO in Australia. 
This research has made use of the NASA Exoplanet Archive, which is operated by the California 
Institute of Technology, under contract with the National Aeronautics and Space Administration 
under the Exoplanet Exploration Program.
Work by IGS and AG was supported by JPL grant 1500811.
AG acknowledges the support from NSF grant AST-1516842.
AG received support from the European Research Council under the European Union's Seventh Framework Programme (FP 7) ERC Grant Agreement n. [321035].
Work by CH was supported by the grant (2017R1A4A1015178) of National Research Foundation of Korea
%



\end{document}